\documentclass[conference]{IEEEtran}
\IEEEoverridecommandlockouts
\usepackage{cite}
\usepackage{amsmath,amssymb,amsfonts}
\usepackage{algorithmic}
\usepackage{graphicx}
\usepackage{textcomp}
\usepackage{xcolor}
\usepackage{tabularx}
\usepackage{hyperref}
\usepackage{array}
\usepackage[utf8]{inputenc}
\usepackage[T1]{fontenc}

\def\BibTeX{{\rm B\kern-.05em{\sc i\kern-.025em b}\kern-.08em
    T\kern-.1667em\lower.7ex\hbox{E}\kern-.125emX}}
\begin{document}

\title{A Virtual Cybersecurity Department for Securing Digital Twins in Water Distribution Systems}

\author{\IEEEauthorblockN{1\textsuperscript{st} Mohammadhossein Homaei}
\IEEEauthorblockA{\textit{Media Engineering Group} \\
\textit{University of Extremadura}\\
C\'aceres, Spain \\
mhomaein@alumnos.unex.es}
\and
\IEEEauthorblockN{2\textsuperscript{nd} Agustin Di Bartolo \\ 3\textsuperscript{rd} Óscar Mogollón-Gutiérrez}
\IEEEauthorblockA{\textit{Media Engineering Group} \\
\textit{University of Extremadura}\\
C\'aceres, Spain\\
\{adibartolo, oscarmg\}@{unex.es}}
\and
\IEEEauthorblockN{4\textsuperscript{th} Fernando Broncano Morgado, \\  5\textsuperscript{th} Pablo García Rodríguez}
\IEEEauthorblockA{\textit{Media Engineering Group} \\
\textit{University of Extremadura}\\
C\'aceres, Spain \\
\{fbroncano, pablogr\}@unex.es}
}

\maketitle

\begin{abstract} Digital twins (DTs) help improve real-time monitoring and decision-making in water distribution systems. However, their connectivity makes them easy targets for cyberattacks such as scanning, denial-of-service (DoS), and unauthorized access. Small and medium-sized enterprises (SMEs) that manage these systems often do not have enough budget or staff to build strong cybersecurity teams. To solve this problem, we present a Virtual Cybersecurity Department (VCD), an affordable and automated framework designed for SMEs. The VCD uses open-source tools like Zabbix for real-time monitoring, Suricata for network intrusion detection, Fail2Ban to block repeated login attempts, and simple firewall settings. To improve threat detection, we also add a machine-learning-based IDS trained on the OD-IDS2022 dataset using an improved ensemble model. This model detects cyber threats such as brute-force attacks, remote code execution (RCE), and network flooding, with 92\% accuracy and fewer false alarms. Our solution gives SMEs a practical and efficient way to secure water systems using low-cost and easy-to-manage tools. \end{abstract}


\section{Introduction} \label{sec:intro}

As water distribution systems become increasingly connected, they face growing cybersecurity risks \cite{Homaei2024, Homaei2022, Abbasi2021, Smith2019Review, Yu2021Survey, Brown2022Survey}. Integrating information technology (IT) and operational technology (OT) has significantly improved efficiency and real-time monitoring, but has also introduced new vulnerabilities. DT technology, which provides virtual replicas of physical systems, further enhances these capabilities by improving operational visibility, predictive maintenance, and decision-making \cite{Tao2018Review, Fuller2020Survey}. However, increased connectivity and intelligence in DTs expand their attack surface, making them vulnerable not only to data leaks but also to threats that could impact public health and infrastructure safety. Cyberattacks such as unauthorized access, data manipulation, or DoS could result in severe incidents like water contamination or system disruptions \cite{Mirchi2020Review}. These attacks can go undetected for long periods, especially in systems without active monitoring. As the number of smart sensors and connected devices grows, the complexity of protecting the infrastructure increases. Thus, robust, automated cybersecurity measures are crucial.

SMEs, which often manage water distribution networks, have serious challenges in cybersecurity because they usually do not have enough money or trained IT staff. Traditional security systems are expensive and need expert teams, so they are not good options for small organizations. To solve this, we propose a VCD, a low-cost and easy-to-use system built with open-source tools. The main tool in the VCD is Zabbix, which gives real-time system monitoring, alerting, and data visualization \cite{Zabbix, Kandasamy2021Survey}. It helps detect technical problems and possible cyber threats in the digital twin environment. To improve detection, we also added a machine-learning-based IDS trained on the OD-IDS2022 dataset. This system can find different types of cyberattacks, such as scanning, brute-force, RCE, and DoS attacks. By combining simple monitoring with advanced machine learning, our framework gives SMEs an effective and affordable way to protect their water systems.

Unlike many existing frameworks, our VCD uniquely combines traditional open-source tools with a customized, explainable machine learning model, all optimized for low-resource environments typical in SME water utilities.

The motivation for this work comes from real needs in the field. Many small and rural water utilities want to use digital twin systems, but they are not ready to face growing cybersecurity risks. Most existing solutions are made for large companies and need expensive tools or professional IT teams. SMEs cannot afford these systems and are left with weak protection. Also, new cyberattacks are becoming smarter and harder to detect with old methods. Our goal is to offer a practical and affordable solution that helps SMEs protect their water infrastructure using tools they can manage themselves. The proposed Virtual Cybersecurity Department gives them a way to use open-source software, automate responses, and improve detection with machine learning—without needing large investments or complex systems.

The remainder of the paper is organized as follows: Section~\ref{sec:related_work} reviews existing research in cybersecurity for DT-based water systems. Section~\ref{sec:proposed_framework} introduces our proposed VCD framework, including system architecture, communication flow, cybersecurity integration, and ML-based IDS. Section~\ref{sec:results} presents the experimental evaluation, including results from the Zabbix-based monitoring and the IDS model performance. Finally, Section~\ref{sec:conclusion} concludes the paper and outlines directions for future work.

\begin{table*}[htbp]
\caption{Recent Work on Cybersecurity in DT-Enabled Water Distribution Systems (Post-2020)}
\scriptsize
\begin{center}
\begin{tabular}{>{\centering\arraybackslash}m{1.5cm}>{\centering\arraybackslash}m{4.5cm}>{\centering\arraybackslash}m{4.5cm}>{\centering\arraybackslash}m{6cm}}
\hline
\textbf{Ref.} & \multicolumn{3}{ c }{\textbf{Focus, Challenges, and Tech. \& Eval.}} \\
\cline{2-4}
 & \textbf{\textit{Focus}} & \textbf{\textit{Challenges}} & \textbf{\textit{Tech. \& Eval.}} \\
\hline
Zhang \emph{et al.} 2021 \cite{Zhang2021} & Attack detection in DT water systems & Distinguish anomalies from normal ops; Integrate IT/OT data & ML anomaly detection; IoT integration; Simulation testbed \\
\hline
Liu \emph{et al.} 2022 \cite{Liu2022} & Secure smart water DTs & Ensuring secure communication & Cryptographic protocols; Anomaly-based IDS; Emulated DT with intrusions \\
\hline
Qi \emph{et al.} 2022 \cite{Qi2022} & Risk assessment in DT networks & Prioritizing vulnerabilities in distributed systems & Sensor fusion; Statistical threat scoring; Risk evaluation scenario analysis \\
\hline
Kumar \emph{et al.} 2023 \cite{Kumar2023} & Mitigate attacks via anomaly+blockchain & Data tampering, traceability & Blockchain for data integrity; ML detection; Experimental deployment \\
\hline
Lin \emph{et al.} 2023 \cite{Lin2023} & IDS using DT correlation (hydraulic/network) & Detect stealthy attacks in normal ops & Hybrid IDS correlating physical \& network metrics; Lab-scale DT with synthetic attacks \\
\hline
\end{tabular}
\label{tab1}
\end{center}
\end{table*}

\section{Related Work} \label{sec:related_work}

\subsection{Cybersecurity in DTs for Water Systems} \label{subsec:cybersecurity_digital_twins}

In recent years, DT technology has become more common in water distribution systems due to its ability to provide real-time monitoring, predictive analytics, and decision support. However, this increased connectivity has also introduced new cybersecurity challenges. DTs, by design, connect multiple physical and digital components, which increases the attack surface for potential cyber threats.

Zhang \emph{et al.}\cite{Zhang2021} presented a machine-learning-based intrusion detection framework for DT-enabled water systems, integrating IoT sensors to detect anomalies in physical and cyber operations. Liu \emph{et al.}\cite{Liu2022} proposed a secure DT architecture using encryption protocols and anomaly-based IDS to protect communication flows between devices and the cloud. Homaei \emph{et al.}~\cite{Homaei2024, Homaei2022} also highlighted the dual role of DTs as both monitoring tools and high-risk targets for attacks, especially in rural water networks.

These studies show that while DTs improve operations, they also require new security solutions that go beyond traditional IT protections.

\subsection{Challenges in DT-based Water Infrastructure} \label{subsec:challenges}

Cybersecurity in water distribution systems faces multiple technical and operational challenges, particularly when DTs are integrated:

\begin{itemize}
    \item Anomaly Detection: DT systems rely on normal behavior patterns to function correctly. However, cyberattacks often mimic legitimate fluctuations (e.g., consumption peaks), making detection difficult without advanced ML techniques.
    \item Scalability and Performance: Real-time monitoring and analysis require high processing power and efficient algorithms, especially as the number of IoT sensors increases.
    \item Legacy and Modern System Integration: Many utilities still use legacy systems that are not easily compatible with modern IoT devices or secure communication protocols, creating interoperability issues.
    \item Network Communication Risks: Protocols used in DTs are sometimes unencrypted or misconfigured, exposing them to packet sniffing, spoofing, or DoS attacks.
    \item Limited Resources in SMEs: Most SMEs lack the IT staff, funding, or training to maintain enterprise-level cybersecurity systems, leaving them especially vulnerable.
    \item Public Safety and Reliability: Failures in cyber-protected DTs could lead to water shortages, contamination, or service disruptions, affecting entire communities.
\end{itemize}

These issues make it clear that new frameworks should be lightweight, scalable, and capable of operating in low-resource environments.

\subsection{Emerging Solutions and Gaps} \label{subsec:existing_solutions}

Recent research has introduced several approaches to improve cybersecurity in DT-enabled water networks. Qi \emph{et al.}\cite{Qi2022} introduced a risk assessment method using sensor fusion and statistical analysis to identify vulnerable components. Kumar \emph{et al.}\cite{Kumar2023} proposed combining blockchain with anomaly detection to increase data traceability and prevent tampering. Lin \emph{et al.}~\cite{Lin2023} focused on hybrid intrusion detection systems that analyze both physical process data and network logs to detect stealth attacks.

Although these methods show progress, they often rely on complex systems or high-performance resources, which may not be suitable for SMEs.

\subsection{Positioning of This Work} \label{subsec:comparison}

In contrast to prior works that require extensive infrastructure or expert personnel, our proposed VCD offers a practical alternative for small and medium-sized enterprises. The VCD uses a combination of lightweight, open-source tools—Zabbix, Suricata, and Fail2Ban—alongside a machine-learning-based IDS trained on the OD-IDS2022 dataset.

Unlike many traditional systems that depend solely on signature-based detection or manual log review, our model integrates real-time monitoring with automated responses and a trained ensemble ML model. This hybrid approach improves detection of advanced threats such as brute-force, RCE, and DoS attacks, making it well-suited for decentralized water systems with limited resources. It also reduces the need for continuous human supervision and simplifies system maintenance, allowing operators to focus on operational tasks rather than complex cybersecurity management.

\section{Proposed Framework} \label{sec:proposed_framework}

This section describes the structure and components of the proposed VCD, a cost-effective monitoring framework for DTs in WDS. The system is designed to help SMEs enhance their operational security through automated, open-source tools. The framework includes four main components: the DT system overview, system architecture and communication flow, cybersecurity integration using Zabbix and Suricata, and a machine learning-based IDS.

\subsection{DT System Overview}
The VCD is built on a Digital Twin platform that integrates real-time data collection, AI-driven analytics, and secure communication. It consists of three main layers: cyber-physical systems (CPS), data management, and predictive analytics.

\begin{figure}
    \centering
    \includegraphics[width=1\linewidth]{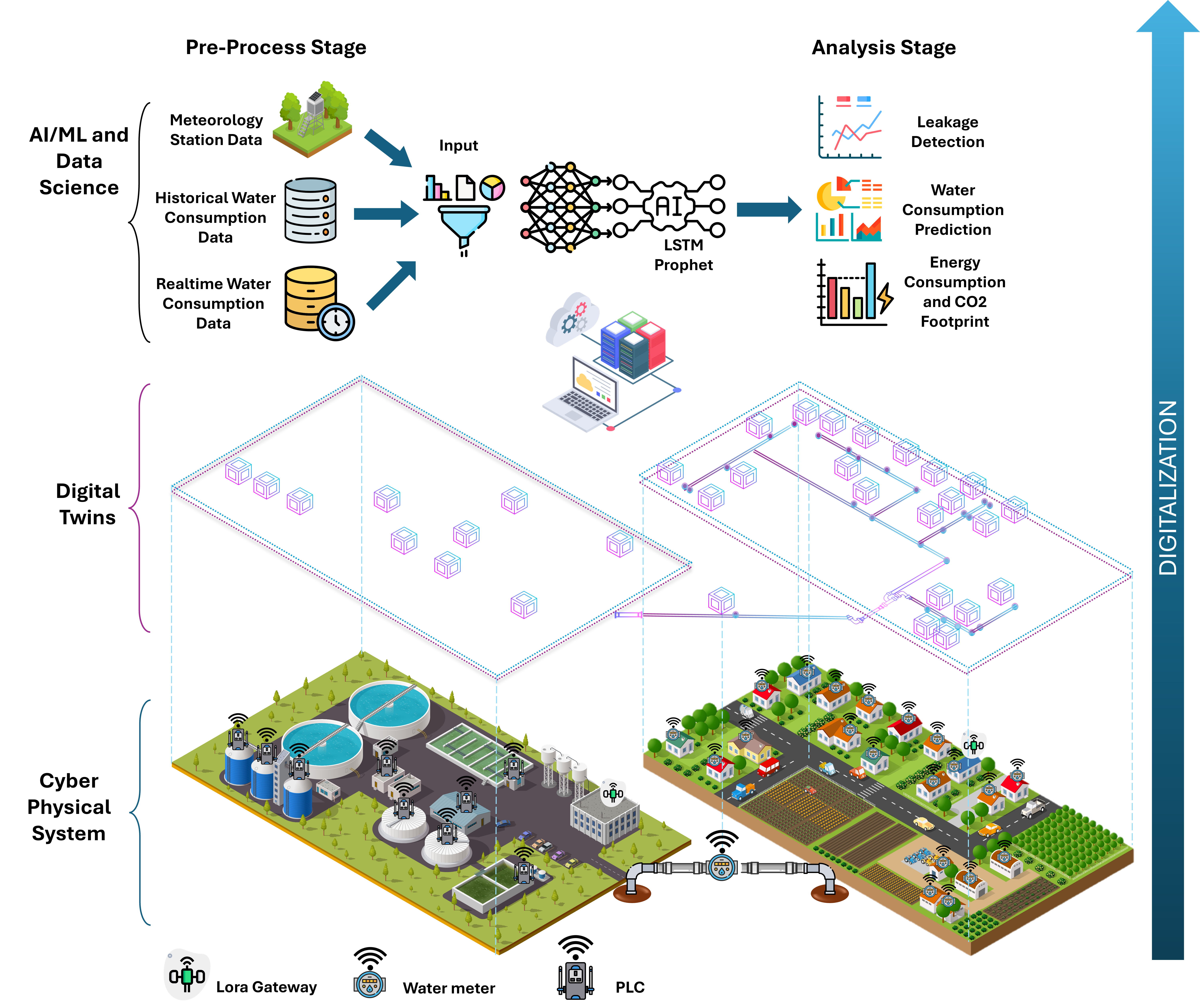}
    \caption{DT platform in the WDS \cite{Homaei2024DTWATER}}
    \label{fig:DTplatform}
\end{figure}

The CPS layer includes sensors, PLCs, and IoT water meters deployed in water treatment facilities and distribution pipelines. These devices collect environmental, operational, and consumption data. The data is transmitted securely using technologies such as LoRaWAN, VPN, and SSH. AI/ML models—including LSTM, Prophet, and LightGBM—are used for water usage forecasting, leakage detection, and energy monitoring. Additionally, GIS tools support spatial analysis and map-based monitoring (Figure \ref{fig:DTplatform})\cite{Homaei2024DTWATER}.

This architecture is designed for rural and small-scale water utilities but is scalable for larger infrastructures. It provides enhanced operational control, cost efficiency, and resource optimization.

\subsection{System Architecture and Communication Flow} \label{subsec:architecture}
The system consists of three key components: edge nodes, secure communication channels, and a central server.

\begin{figure}
    \centering
    \includegraphics[width=0.65\linewidth]{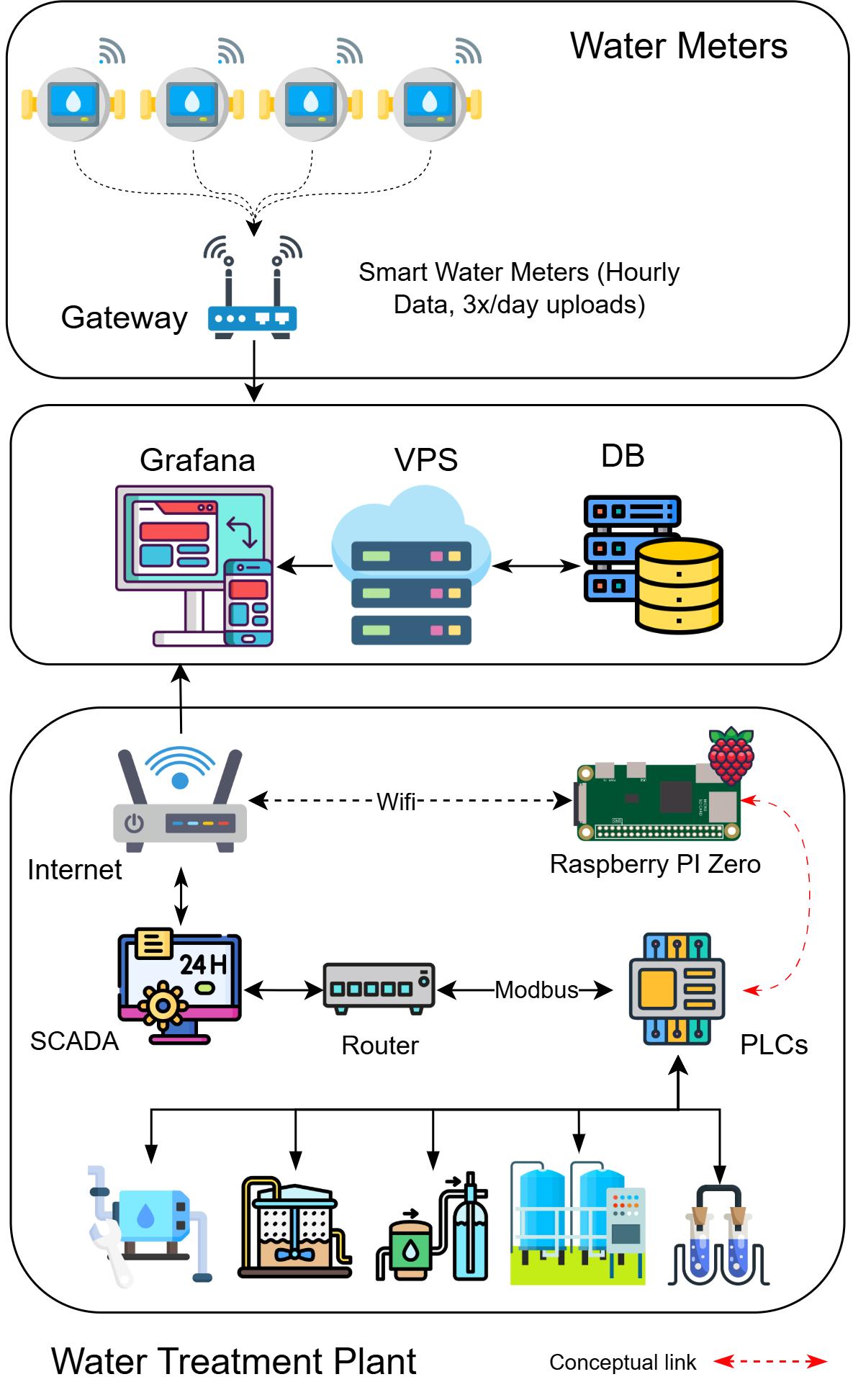}
    \caption{Deployment of Zabbix proxies on Raspberry Pi devices for real-time data collection from IoT meters and SCADA systems.}
    \label{fig:Pi_collect_data}
\end{figure}

\begin{itemize}
    \item Edge nodes include Raspberry Pi devices equipped with Zabbix proxies, IoT meters, SCADA units, and PLCs. These nodes are strategically placed at water plants and administrative locations to ensure complete visibility (Figure \ref{fig:Pi_collect_data}).
    \item Secure communication is established using VPN tunnels, SSH protocols, and LoRaWAN networks. This ensures that data from the field devices reaches the central server with integrity and confidentiality. Zabbix continuously monitors the stability and quality of these connections.
    \item The central server, hosted on a Virtual Private Server (VPS), aggregates all incoming data. It runs Zabbix for real-time monitoring, Suricata for intrusion detection, and Fail2Ban for automated IP blocking. The server can optionally connect to cloud platforms like AWS or Azure for data storage and computational scalability.
\end{itemize}

\begin{figure*}[htb]
    \centering
    \includegraphics[width=\textwidth]{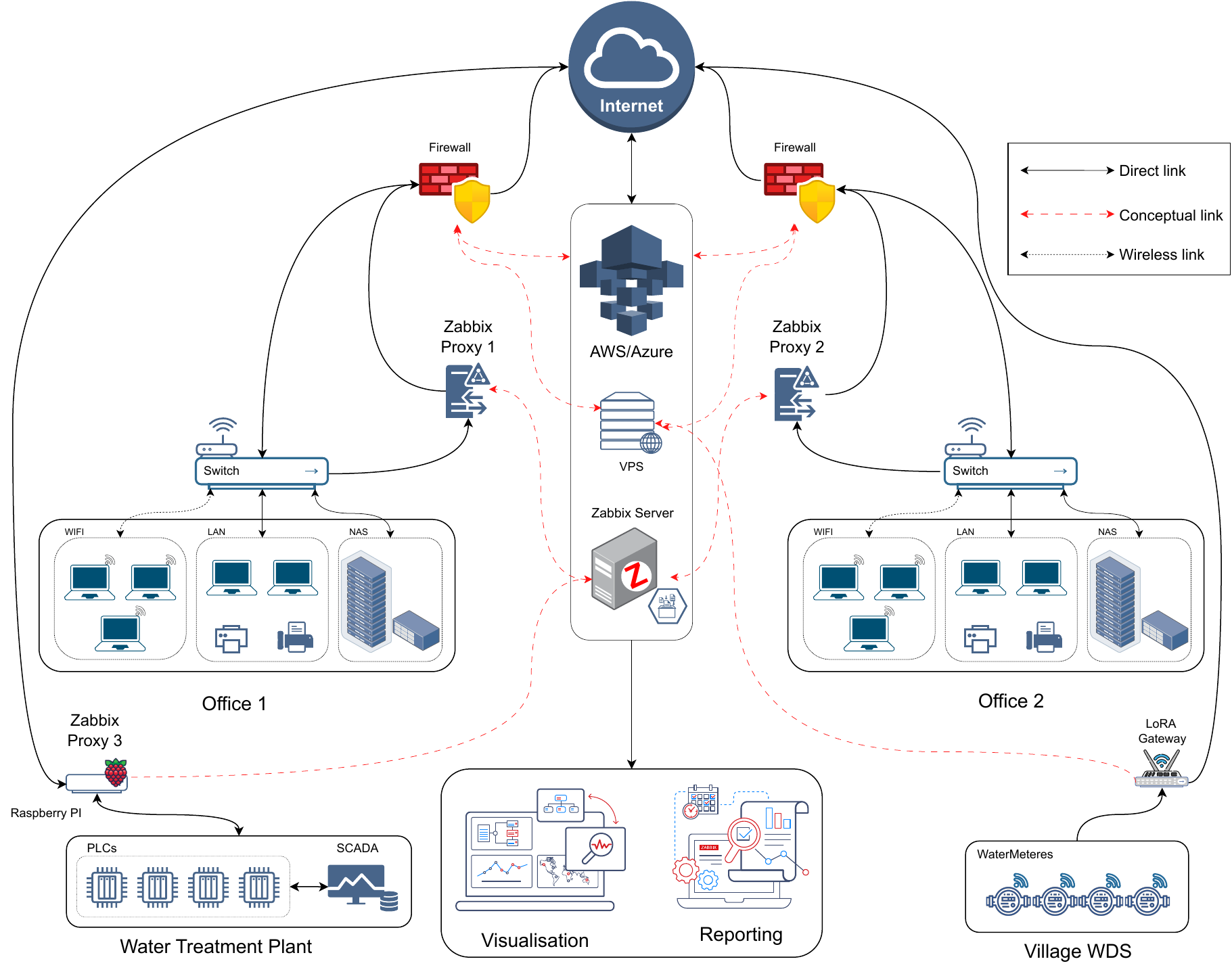}
    \caption{VCD architecture with Zabbix and ML-based IDS for DT-enabled SME water systems}
    \label{fig:Zabbix}
\end{figure*}

Figure~\ref{fig:Zabbix} presents the VCD architecture, highlighting the placement of Zabbix and the ML-based IDS modules.

\subsection{Cybersecurity Integration with Zabbix and Suricata}
The core of the cybersecurity layer is Zabbix, which provides data collection, visualization, and alerting functionalities. It monitors metrics such as network traffic, CPU load, memory usage, and failed login attempts. Zabbix is integrated with Suricata, an open-source IDS that inspects network packets and detects threats like port scanning, brute-force logins, and unusual data flows. Suricata's alerts are visualized in the Zabbix dashboard. Fail2Ban complements the system by monitoring authentication logs. It automatically bans IP addresses that exceed a defined number of failed login attempts. This combination of tools ensures multi-layered protection against a wide range of attacks while remaining lightweight and suitable for resource-constrained environments.

\subsection{AI/ML-Based Intrusion Detection System} \label{subsec:ml_ids}

As part of the proposed framework, we developed a machine learning-based IDS to improve the detection of cyber threats in smart water networks. This IDS is trained on the OD-IDS2022 dataset, which provides 1,031,916 labeled samples \cite{Patel2023}. Each sample contains 82 features representing flow-based network data, including IP addresses, port numbers, protocol types, packet lengths, time durations, and flag behaviors. These records include normal traffic and 29 attack types such as DoS, brute force, SQL injection, RCE, hijacking, and reconnaissance. To simplify classification and reduce overfitting, we grouped the 29 attack classes into seven general categories, listed in Table~\ref{tab:attack_groups}. This grouping keeps the detection meaningful while making the machine learning models easier to train and evaluate.

\begin{table}[ht]
\tiny
\caption{7-Group Attack Categorization}
\centering
\begin{tabular}{ll}
\hline
\textbf{Group} & \textbf{Includes} \\
\hline
BENIGN      & BENIGN \\
DOS         & DoS Hulk, Slowhttptest, GoldenEye, Slowloris, DDoS-* \\
BRUTEFORCE  & Bruteforce-Web, Bruteforce-XSS, FTP/SSH-Patator, Web Brute Force \\
INJECTION   & SQL/LDAP/SIP Injection, Web SQL Injection \\
HIJACKING   & MITM, Hijacking \\
RCE         & RFI, Exploit, Cmd Injection, Upload, Backdoor \\
OTHER       & Infiltration, Bot, PortScan, Web XSS \\
\hline
\end{tabular}
\label{tab:attack_groups}
\end{table}

We proposed and implemented five machine learning models as part of the IDS component. All models use the same preprocessing pipeline: label encoding, numerical feature extraction, mutual information for feature selection, data normalization, and oversampling with SMOTE to balance class distribution.

\subsubsection{Random Forest Classifier}

The Random Forest (RF) model builds many decision trees from random subsets of the training data. Each tree gives a class prediction, and the final result is selected by majority voting. This is expressed in Equation~\ref{eq:rf}.

\begin{equation}
\hat{y} = \mathrm{mode} \left( h_1(x), h_2(x), \dots, h_T(x) \right)
\label{eq:rf}
\end{equation}

where \( h_t(x) \) is the prediction of tree \( t \), and \( T \) is the total number of trees.

\subsubsection{Tuned LightGBM Classifier}

LightGBM is a gradient boosting algorithm that builds trees sequentially to minimize prediction errors. It grows trees leaf-wise and uses a loss function with regularization, as shown in Equation~\ref{eq:lgbm}.

\begin{equation}
\mathcal{L}^{(t)} = \sum_{i=1}^n L(y_i, \hat{y}_i^{(t-1)} + f_t(x_i)) + \Omega(f_t)
\label{eq:lgbm}
\end{equation}

Here, \( L \) is the loss function, \( \hat{y}_i^{(t-1)} \) is the previous prediction, \( f_t \) is the new decision tree, and \( \Omega \) is the regularization term.

\subsubsection{Improved Ensemble Model (v1)}

This model combines three base classifiers—Random Forest, LightGBM, and a Multi-layer Perceptron (MLP)—into a soft voting ensemble. It averages the class probabilities from each model and selects the class with the highest average score, as shown in Equation~\ref{eq:ensemble_v1}.

\begin{equation}
\hat{y} = \arg \max_c \left( \frac{1}{M} \sum_{m=1}^{M} P_m(c \mid x) \right)
\label{eq:ensemble_v1}
\end{equation}

where \( P_m(c \mid x) \) is the probability of class \( c \) predicted by model \( m \), and \( M \) is the number of models.

\subsubsection{Weighted Ensemble with Feature Engineering}

This model improves ensemble voting by assigning custom weights to each classifier and using new engineered features like packet length ratios and size variations. The prediction formula with weights is given in Equation~\ref{eq:weighted_ensemble}.

\begin{equation}
\hat{y} = \arg \max_c \left( \sum_{m=1}^{M} w_m \cdot P_m(c \mid x) \right)
\label{eq:weighted_ensemble}
\end{equation}

where \( w_m \) is the weight assigned to model \( m \), and \( \sum w_m = 1 \).

\subsubsection{Improved Ensemble (v2)}

The final and most optimized model uses the same weighted voting as in Equation~\ref{eq:weighted_ensemble}, but with improved components. These include:
\begin{itemize}
    \item A deeper MLP with 3 hidden layers (256, 128, 64) and ReLU activation
    \item A tuned LightGBM with max depth = 10, 64 leaves, and learning rate = 0.05
    \item A larger Random Forest with 150 trees and class-balanced weighting
\end{itemize}

The ensemble weights are selected based on validation scores to ensure balanced detection across all classes, especially minority attacks like RCE and Hijacking.

\noindent\textit{Note:} To improve model transparency, we use SHAP (SHapley Additive exPlanations), a method from cooperative game theory that attributes prediction changes to individual features. The SHAP value for a feature \( i \) is calculated using:

\begin{equation}
\phi_i = \sum_{S \subseteq F \setminus \{i\}} \frac{|S|!(|F| - |S| - 1)!}{|F|!} \left[ f(S \cup \{i\}) - f(S) \right]
\label{eq:shap}
\end{equation}

Here, \( F \) is the full feature set, \( S \) is a subset of features excluding \( i \), and \( f \) is the prediction function. SHAP values explain how much each feature contributes to the final prediction, helping operators understand the decision process of the IDS.

In summary, the AI-based IDS module strengthens the proposed framework by enabling real-time detection of various cyber threats using interpretable and resource-efficient machine learning models. In the following section, we present the experimental evaluation and performance results of the IDS models, along with the integration of Zabbix for continuous monitoring in the digital twin environment.

\section{Experimental Evaluation and Model Performance} \label{sec:results}

This section presents the experimental evaluation of the proposed VCD for water systems. The validation includes two parts: real-time monitoring results using Zabbix, and the performance of machine learning models for intrusion detection.

\subsection{Monitoring Setup and Attack Simulation}

The VCD was tested in a hybrid digital twin setup. Zabbix server was installed on a VPS, and several Raspberry Pi devices were installed in field locations like water plants and offices. These Raspberry Pis worked as Zabbix proxies and collected logs from IoT meters, PLCs, and SCADA systems.

\begin{figure}[htb]
    \centering
    \includegraphics[width=1\linewidth]{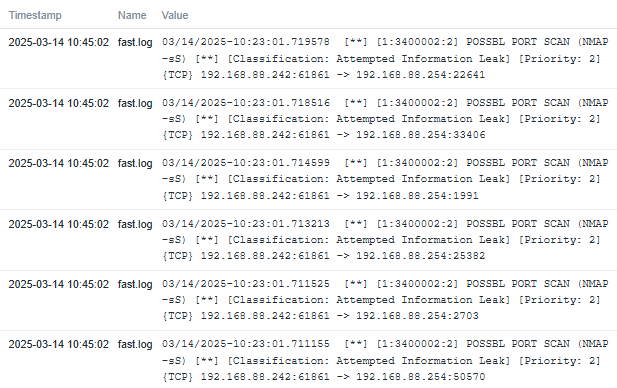}
    \caption{logging attempt to the servers}
    \label{fig:login}
\end{figure}

\begin{figure}[htb]
    \centering
    \includegraphics[width=1\linewidth]{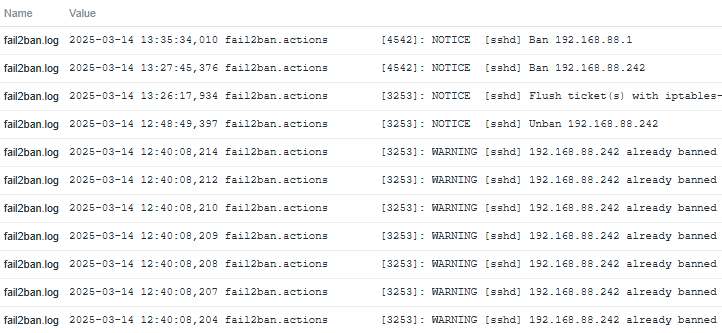}
    \caption{Fail2Ban logs showing IP bans triggered by repeated failed SSH login attempts}
    \label{fig:Fail2Ban}
\end{figure}

\begin{figure}[htb]
    \centering
    \includegraphics[width=0.95\linewidth]{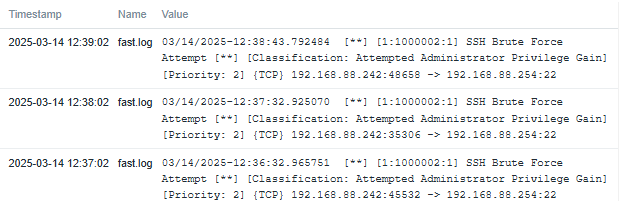}
    \caption{Suricata alerts for SSH brute-force attack attempts showing repeated unauthorized access to port 22}
    \label{fig:SSHSuricata}
\end{figure}

To test the system, three types of cyberattacks were performed:

\begin{itemize}
    \item \textit{Nmap Scan (Reconnaissance)}: A stealth scan was launched using Nmap to find open ports. Suricata detected this scan and sent alerts to Zabbix. The traffic pattern showed abnormal packet behavior (Figure~\ref{fig:login}).
    
    \item \textit{Brute Force (Hydra + SSH)}: An SSH brute-force attack was simulated using Hydra. Zabbix recorded many failed logins and increased CPU usage. Suricata also detected frequent access to port 22. Fail2Ban blocked the attacker's IP after too many failed attempts, as shown in Figures(~\ref{fig:Fail2Ban}, \ref{fig:SSHSuricata}).
    
    \item \textit{DoS (hping3)}: A SYN flood attack was done using hping3. It caused high CPU and memory usage. Zabbix showed this unusual behavior and created alerts, even when the logs were not clear.
\end{itemize}

These tests showed that the system can detect and respond to real cyberattacks using simple and open-source tools.

\subsection{Monitoring Indicators}

\begin{figure}[htb]
    \centering
    \includegraphics[width=1\linewidth]{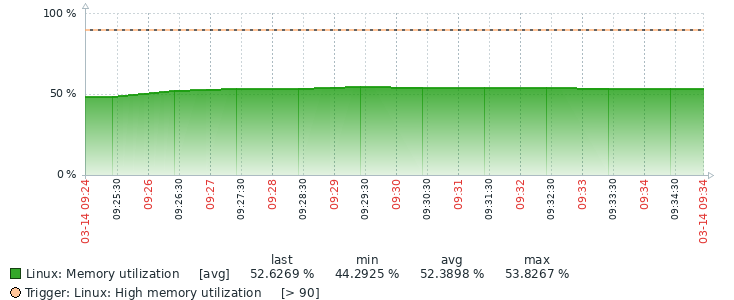}
    \caption{Memory usage monitoring under DDoS Attack}
    \label{fig:memddos}
\end{figure}

\begin{figure}[htb]
    \centering
    \includegraphics[width=1\linewidth]{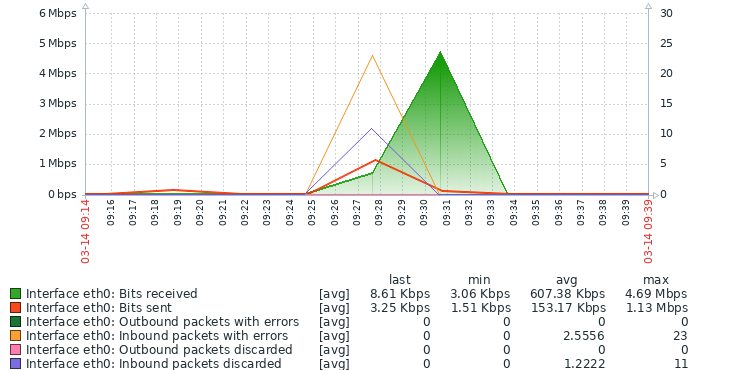}
    \caption{Network monitoring under DDoS Attack}
    \label{fig:netddos}
\end{figure}

Several indicators were collected from Zabbix to check the system behavior:

\begin{itemize}
    \item \textit{CPU and Memory Usage}: These increased during DoS attack and helped to detect it (Figure~\ref{fig:memddos}).
    \item \textit{Network Traffic (Upload/Download)}: Abnormal traffic helped detect Nmap and DoS attacks (Figure~\ref{fig:netddos}).
    \item \textit{Dropped and Malformed Packets}: These increased during the flood attack.
    \item \textit{Failed Login Attempts}: Zabbix tracked this for brute force detection, and Fail2Ban blocked the IP.
    \item \textit{Suricata Alerts}: Number of alerts helped show which attack was happening.
    \item \textit{Alert Time}: The system created alerts in a few seconds after the attack started.
\end{itemize}

\subsection{Machine Learning IDS Evaluation}

Besides traditional detection, five machine learning models were tested using the OD-IDS2022 dataset. The goal was to classify seven types of network traffic, including attacks like RCE, hijacking, and injection.

Table~\ref{tab:model_comparison} shows the performance of each model. The best model was the improved ensemble (v2), which used LightGBM, Random Forest, and MLP together.

\begin{table}[ht]
\tiny
\caption{Comparison of IDS Models for 7-Class Categorization (OD-IDS2022 Dataset)}
\centering
\begin{tabular}{lcccccc}
\hline
\textbf{Model} & \textbf{Acc.} & \textbf{Macro F1} & \textbf{RCE F1} & \textbf{HIJACK Rec.} & \textbf{Explainable} & \textbf{Gran.} \\
\hline
Random Forest        & 77.0\% & 0.47   & 0.37  & 0.54  & \checkmark~SHAP & 30+ \\
LightGBM (Tuned)     & 82.2\% & 0.714  & 0.526 & 0.609 & -- (addable)    & 7   \\
Improved Ens. (v1)   & 80.4\% & 0.6645 & 0.55  & 0.73  & \checkmark~SHAP & 7   \\
Weighted Ens. + FE   & 80.2\% & 0.66   & 0.54  & 0.76  & \checkmark~SHAP & 7   \\
Improved Ens. (v2)   & 92.0\% & 0.88   & 0.86  & 0.87  & \checkmark~SHAP & 7   \\
\hline
\end{tabular}
\label{tab:model_comparison}
\end{table}

\begin{figure}
    \centering
    \includegraphics[width=1\linewidth]{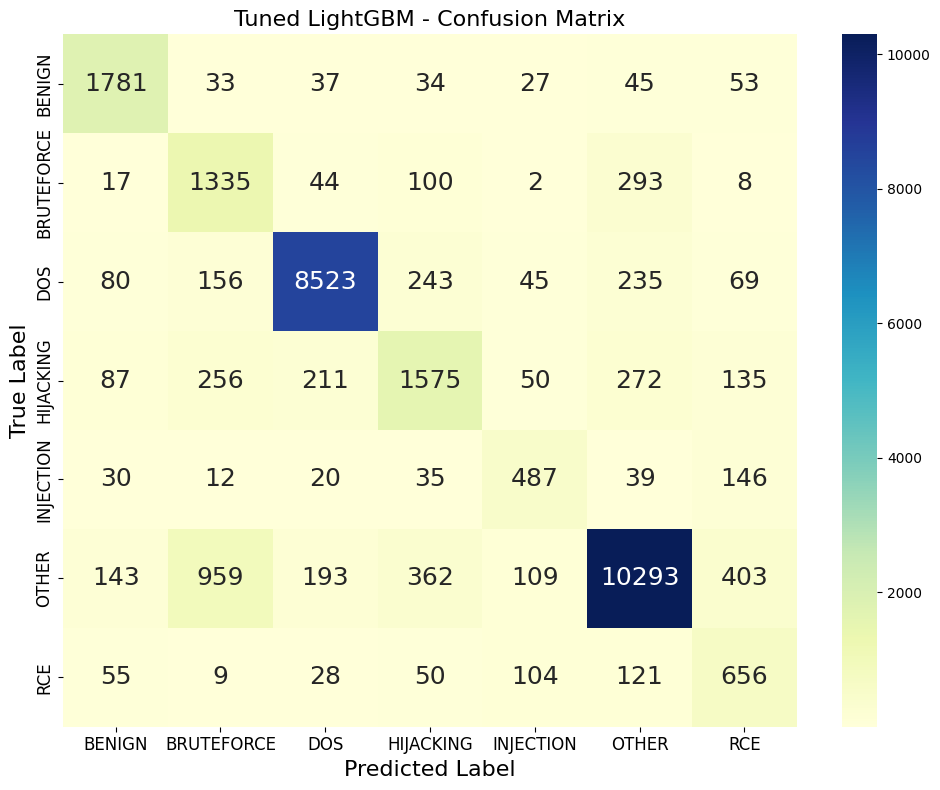}
    \caption{Confusion matrix showing class-wise prediction performance across 7 traffic categories.}
    \label{fig:enter-label}
\end{figure}

Table~\ref{tab:classification_report} shows the full report for the best model. It gives good results in all classes, including small ones like injection. Figure~\ref{fig:enter-label} shows the confusion matrix.

\begin{table}[ht]
\caption{Ensemble Model Classification Report}
\centering
\begin{tabular}{lcccc}
\hline
\textbf{Class} & \textbf{Precision} & \textbf{Recall} & \textbf{F1-score} & \textbf{Support} \\
\hline
BENIGN      & 0.91 & 0.93 & 0.92 & 2024 \\
BRUTEFORCE  & 0.87 & 0.85 & 0.86 & 2377 \\
DOS         & 0.96 & 0.97 & 0.96 & 6463 \\
HIJACKING   & 0.84 & 0.87 & 0.85 & 2475 \\
INJECTION   & 0.82 & 0.78 & 0.80 & 220  \\
OTHER       & 0.92 & 0.93 & 0.93 & 14629 \\
RCE         & 0.85 & 0.88 & 0.86 & 1812 \\
\hline
Accuracy &       &       & 0.92 & 30000 \\
Macro Avg & 0.88 & 0.89 & 0.88 & 30000 \\
Weighted Avg & 0.92 & 0.92 & 0.92 & 30000 \\
\hline
\end{tabular}
\label{tab:classification_report}
\end{table}

This experiment confirms that the proposed VCD can detect and respond to cyberattacks in real time using both rule-based and AI-based tools. It works well even in small water systems with low-cost hardware.

\section{Conclusion and Future Work} \label{sec:conclusion}

This study proposed a VCD to improve the cybersecurity of DTs used in water networks, with a focus on SMEs. The solution combines free tools: Zabbix for live monitoring, Suricata as an IDS, and Fail2Ban to block repeated login attempts. Zabbix proxies were installed on Raspberry Pi units to collect data from SCADA, PLCs, and IoT sensors. We tested the system with simulated attacks (port scanning, brute-force on SSH, and DoS), and it responded correctly with alerts, log collection, and IP blocking.

The IDS part was developed using the OD-IDS2022 dataset (over one million records with 29 classes). We simplified the task by grouping the classes into 7 attack types. We tested five ML models, and the final version (v2) used a combination of RF, LGBM, and MLP with SHAP for explainability. This model gave the best results for detecting attacks like RCE, hijacking, and injection. The full framework is low-cost, supports real-time detection, and works well for small organizations without advanced computing systems.

For future work, we will explore LLMs to improve detection accuracy, reduce false positives, and classify threats more precisely. We also plan to integrate blockchain to protect data integrity and support trusted operations. These upgrades aim to create smarter and more secure water management systems.

\section*{Acknowledgment}
This initiative is carried out within the framework of the funds from the Recovery, Transformation, and Resilience Plan, financed by the European Union (Next Generation) – National Institute of Cybersecurity (INCIBE), as part of project C107/23:
"Artificial Intelligence Applied to Cybersecurity in Critical Water and Sanitation Infrastructures."


\end{document}